\documentclass[journal,11pt,draftclsnofoot,onecolumn]{IEEEtran}
\usepackage {graphicx} 
\usepackage {subfigure}
\usepackage {amsthm}
\usepackage[fleqn] {amsmath}
\usepackage {amsfonts, amssymb}
\usepackage{multirow}

\title{Using TV Receiver Information to Increase Cognitive White Space Spectrum}
\author{Brage~Ellings\ae ter,~\IEEEmembership{Student Member,~IEEE,}
				Hemdan~Bezabih, Josef~Noll and
				Torleiv Maseng,~\IEEEmembership{Member,~IEEE}}
\begin{document}
\maketitle

\begin{abstract}
In this paper we investigate the usage of cognitive radio devices within the service area of TV broadcast stations. Until now the main approach for a cognitive radio to operate in the TV bands has been to register TV broadcast stations locations and thus protecting the broadcast stations service area. Through information about TV receivers location, we show that a cognitive radio should be able to operate within this service area without causing harmful interference to the TV receivers as defined by Ofcom and FCC. We provide simulations based on real statistics from Norway that show that especially in rural areas TV receiver registration can provide a substantial gain in terms of exploitable frequencies for a cognitive radio.
\end{abstract}


\section{Introduction}
In the last decade demand for wireless services has exploded. According to Cisco, mobile data traffic is expected to have a 26-fold increase from 2010 to 2015 \cite{cisco}. It might seem impossible to increase the cellular capacity by such a factor, but in fact the cellular capacity has increased substantially if we compare 3G/4G technology to the earliest analog technologies. Increasing cellular capacity can essentially be done by three factors: (i) increase in spectral efficiency (b/s/Hz), (ii) increase in available bandwidth and (iii) increase in number of base stations.

Increasing spectral efficiency can be done by different techniques. The most essential is to use good error correcting codes to approach Shannon's capacity. However, with turbo codes and LDPC codes we are at only a small delta away from the channel capacity and the hope for increasing spectral efficiency by a significant factor due to coding is almost non-existing. Other ways of increasing spectral efficiency, such as MIMO technology and base station cooperation and multiuser detection, are to some extent present in the specifications of future cellular technologies such as LTE Advanced \cite{4907410}\cite{3gpp1}. In addition, the complexity of the wireless devices increases significantly when these techniques are implemented and some fundamental properties (such as necessary antenna spacing for MIMO gain) limits the gains of these techniques.

If we look at the evolution of mobile data traffic over the last 20 years, the factor that has increased capacity the most is an increased number of base stations (and thus cell-size reduction).
Today, femto cells are the most appealing technology to provuide high bit rates because femto cells can give high bit rates essentially due to the small cell size. However, decreasing cell sizes leads to more handovers, which complicates the system, and one can only decrease the cell-size so much until inter-cell interference limits performance.

The last factor is bandwidth. For instance, LTE Advanced is suppose to achieve the 4G speeds of maximum 1 Gb/s/cell in the downlink by aggregating a total of 100 MHz of bandwidth (in addition to $8\times 8$ MIMO) \cite{3gpp1}. The problem is that in an actually deployed system, 100 MHz is hard to come by. In fact, by the spectrum regulations done in most countries, no single operator can expect to be able to aggregate 100 MHz. Spectrum is limited due to the small amount of spectrum that have desirable attenuation properties. 

When TV broadcasting systems switched from analog to digital broadcasting, roughly 100 MHz was freed up depending on the country. This part of the spectrum has become known as the digital dividend, and specific re-allocation of this spectrum to other services is a current process in many countries. Due to the high value of spectrum, it is likely that only the powerful operators will be able to buy the rights to these frequencies.

As the need for spectrum for high capacity services became apparent, cognitive radio emerged as a hot research topic \cite{mitola}. The basic idea of cognitive radio is that a band allocated to a certain wireless service will not always be used. Thus, if an unlicensed user (secondary user) could be able to detect these spectral holes it should be able to utilize them without causing harmful interference to the licensed user (primary user). Over the years more and more research has been done in this area, especially in the areas of sensing and simultaneous transmission between primary and secondary users. Although a number of fundamental limits on cognitive radio has been found, cognitive radio is still in its inception with standards such as IEEE 802.22 \cite{wran} just emerging.

One part of the spectrum particularly suited for cognitive radio is the frequencies allocated for terrestrial Digital Video Broadcasting (DVB-T). This is due to the reuse factor used in DVB-T systems, which leads a portion of the allocated spectrum unused at any given location. The under utilization of these frequencies has been verified both by the FCC \cite{fcc5} and Berlemann and Mangfold in Switzerland \cite{BerlemannMangold2009}. These unused frequencies are known as the TV white spaces. Many researchers have pushed hard to allow cognitive radio technology to operate in the TV white spaces, and recently the FCC has allowed cognitive radios to operate in the white spaces, given they do not create harmful interference to the TV receivers. The current criteria for transmission for a cognitive radio is that the radio must be able to look up its location in a database and transmit under the power constraint given for that frequency in that area \cite{FCC_2010}. The criteria for transmission is primarily based on the TV broadcast station location and cognitive radio location. However, in many countries it is still up to the license holders of the spectrum to decide if cognitive radio technology can utilize the bands or not.

Only registering the TV broadcast stations leads to a significant overprotection of certain areas, areas where there are no TV receivers. But it also reduces potential revenue of the TV broadcast corporations in many countries. For instance, in Norway NTV holds the rights for transmission on the DVB-T bands. They can offer to lease the white spaces to other users that needs spectrum, but if they can only provide a small amount of spectrum, then the revenue will be low. If they could lease out a high amount of spectrum and still guarantee protection to the TV receivers, revenue could be high.

In \cite{hemdan} we introduced the concept of registering TV receiver locations as an alternative to TV broadcast station registration. Many so-called smart TVs are already equipped with an Internet connection, so the thought of registering TV receivers is not as far fetched as it would have seemed 10 years ago. We provided some initial simulation results from Norway where household locations were used as TV receiver locations with location resolution of 1 km. These initial results showed that TV receiver information has a clear advantage over TV broadcast station information and forms the basis for this paper. In this paper we extend the previous results by obtaining more accurate TV receiver information (resolution of 100 m compared to 1 km) and we also compare amount of available spectrum by only knowing TV receiver locations, knowing the decodable channels at a TV receiver and having real-time knowledge of TV viewings. 

We define \textit{white spaces} in the conventional way as \textit{spectrum that is unused by the TV broadcast station in the area due to the reuse factor in DVB-T systems}. We introduce \textit{gray spaces} as \textit{additional spectrum that is available given TV receiver information}. To estimate the amount of gray space we define the protection area of a TV transmitter based on tolerable interference levels for a TV receiver given by the FCC and Ofcom and a standard propagation model (Okumura-Hata model). We assume two different types of wireless devices as cognitive radios. The first is a fixed 4 W radio, which is to illustrate an IEEE 802.22 base station. The second is a mobile 100 mW radio, which is thought of as a hand held device, such as a mobile phone, laptop or other small scale radio device.

We also define three knowledge levels of TV receiver information that yields different amount of gray space:
\begin{itemize}
\item Knowledge Level 1 (KL1): Only TV receiver locations. \\
\item Knowledge Level 2 (KL2): TV receiver locations and TV channel subscriptions.\\
\item Knowledge Level 3 (KL3): TV receiver locations and current MUX usage if each TV receiver. 
\end{itemize}
For these knowledge levels we provide simulation results based on real location information from Norway. The main findings is that by registering TV receivers, significant amount of frequencies can be utilized by cognitive radios in addition to white spaces, especially for power limited devices.

The rest of this paper is organized as follows: Section \ref{sec:2} is regarding unlicensed operation in the TV white spaces, Section \ref{sec:tvrx-reg} is on TV receiver registration, Section \ref{sec:cr-param} defines the cognitive radio parameters, Section \ref{sec:sim-setup} and \ref{sec:sim-res} presents the simulation setup and results, Section \ref{sec:discussion} discusses the results and Section \ref{sec:conclusion} concludes the paper.

\section{Unlicensed Operation in the TV White Spaces}
\label{sec:2}
\begin{figure}
\centering
\includegraphics[width = 0.5\columnwidth]{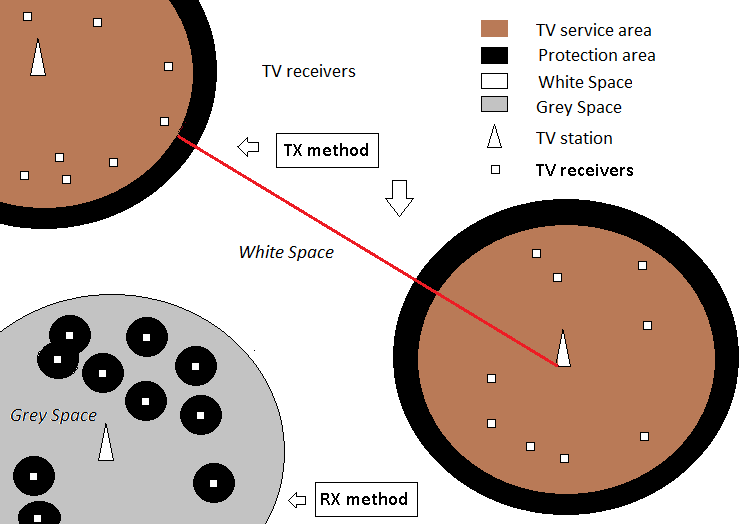}
\caption{The concept of white space and gray space. The three circles represent TV coverage areas where the broadcast stations uses the same frequency. By only registering the TV broadcast stations location's, this frequency can only be used by a CR device outside the protection area of these coverage areas (white space). By registering the TV receivers and protecting these receivers, a CR device can utilize the same frequency as the TV broadcast station transmits on inside the coverage area (gray space).}
\label{fig:concept}
\end{figure}

As mentioned above, white spaces occur due to the reuse factor used in DVB-T systems. Or more specifically because there is a necessary separation distance between two TV broadcast stations using the same frequencies to avoid interference. This is illustrated in Figure \ref{fig:concept}. It is possible for other wireless devices to operate in the frequency bands used by TV broadcast stations sufficiently far away with lower power than the TV broadcast stations, without affecting the reception of the TV receivers.

\subsection{Amount of White Space}
There has been a lot of work done trying to estimate the amount of unused frequencies in the band allocated for DVB-T. The amount of white space found depends on location, the transmit power of the CR device and the restrictions put on what channels to use \cite{WS_UK,Mubaraq_1how,Scott06measuringthe}. E.g. \cite{surveyTV} estimated the average amount of white space spectrum in any location in the UK to be 150 MHz. The calculation considers low-power devices with a power of 100 mW and were there are no constrains on adjacent channel usage. By also taking into account restrictions on adjacent channels the amount of available spectrum decreases to an average of 30 MHz. In the US, Snider estimated that the average amount of white space per person is 214 MHz\cite{snider}. This was based on estimates from the FCC that on the average each person can receive 13.3 MUXs of a total of 49 MUXs, where each MUX is 6 MHz.

However, it is difficult to put an exact number on the amount of white space as different measurement techniques and aspects considered yield different amounts. In \cite{harrison} it is pointed out that two reported amounts differed by as much as a factor of 2, due to different the different restrictions considered.

\subsection{Accessing White Space}
When operating on a secondary basis in licensed bands the main consideration is the licensed users. In the TV band this is mainly the TV receivers and wireless microphones. Both in the US and UK different methods have considered with regards to accessing and utilizing white space. Currently, the approach approved by the FCC is the use of a geo-location database. This database registers TV transmitters and their service area, and in addition other protected devices like cable head ends and locations where wireless microphones are used are registered. In this paper this database is called the TX database. This information ensures that the TV receivers within the service area are protected; the channels used within the service area are set as occupied and are not available for a CR device. 

This approach requires the unlicensed device to supply its geographical coordinates to the database to retrieve those channels that are unused at the device's current location. Although this requires the CR device to have some means of communicating with the database, due to issues related to sensing, such as the hidden node problem, it is the approach agreed upon by regulators both in the US and UK \cite{FCC_2010,ofcom_2009}.

\subsection{Operation Rules}
Once a CR device has supplied the TX database with its location and received the set of available frequencies, there are still some operational requirements. These are primarily requirements on the co-channel and adjacent channel interference levels.

The maximum level of interference a TV receiver can handle is defined by the carrier-to-interference ratio, $C/I$. The $C/I$ levels given below by regulators defines \textit{how much lower} the CR signal strength must be compared to the TV signal at the TV receiver. Ofcom has set the $C/I$ level for co-channels interference as 33 dB and -17 dB for adjacent channel interference. In their calculations they have taken into account other co-channel and adjacent channel interference sources that may operating simultaneously, and is therefore a conservative approach \cite{ofcom_2009}. FCC operate with a $C/I$ level of 23 dB for co-channel interference and -26/-28 dB for upper and lower adjacent channel interference respectively. The technical parameters defined by Ofcom and FCC are summarized in Table \ref{tab:1}.

\section{TV Receiver Registration}
\label{sec:tvrx-reg}
By the current rule of a TX database, all possible user locations within the service area of a given TV transmitter are protected. This method is conservative regarding utilizing frequencies as it protects all locations within an area without considering if there is anyone to protect. TV receivers are not located everywhere within a service area, thus making this method ineffective especially in rural areas. Examples of such scenarios are given below:
\begin{itemize}
\item Of the total population only a portion uses the digital terrestrial television. But using the TX database those who are not using the TV service are also protected.
\item There may be households within the service area that are receiving signals from other TV stations using a different frequency. Using the TX database gives no options of adjusting the protection to fit certain cases.
\item TV coverage areas may overlap which makes it hard to make a statement regarding which channels are used at a location. Using the TX database may lead to locations intersecting with several TV service areas, which will increase the overprotection. 
\item Signal strength of the TV transmitter will vary within a service area; there are locations that do not have access to the service or were the probability of service is low because of terrain and structure. There is no need to protect those areas within the service area. 
\end{itemize}
The common factor for these issues is that registration of TV receiver location together with terrain knowledge would to a large extent solve these issues.

One aspect we have to take into account with this method is how the registration of TV receiver location can be done. One solution can be to provide the location information when buying the TV. In some countries, e.g. Norway, this is already done. Not to enable a TV receiver location database, but due to a mandatory license needed to watch TV. The second, and perhaps most appropriate option, is for the TV user to register the TV receiver location online. The incentive to the user is of course that his signal reception will be better protected against cognitive radio.

One problem is portable TV devices. If such a device has an IP connection it can provide updates on its location. But as the majority of TV receiver locations are fixed receivers, we neglect this issue in this paper.

\subsection{TV Receiver Protection}
With knowledge of the TV receivers we no longer have to protect the whole service area of a TV broadcast station from unlicensed transmission. Instead we are able to construct protected areas around the known TV receivers. The spectrum made available to a CR device by this method is defined as \textit{gray space}.

With knowledge of the TV broadcast station location and TV receiver location, one can estimate the signal strength of the TV signal at the TV receiver. Using the known criteria for carrier-to-interference level given in Table \ref{tab:1}, one can estimate the protection radius around the TV receiver. However, the uncertainty in the estimated signal strength may result in an under protection of the TV receiver. The conservative approach would be to assume the lowest possible signal strength that a receiver can have while still be regarded as belonging to the broadcast station's service area. According to Ofcom the minimum field strength needed at the TV receiver is 50 dB$\mu$V/m, and we will assume all TV receivers operate at this level in the rest of the paper.

The accuracy of the TV receiver location will also be an issue. If we assume the user can provide GPS coordinates of the location we have to have some margin of error due to inaccuracy in the GPS device. Also, the TV receiver antenna is unlikely to be located right next to the TV, thus additional margin of error has to be assumed. Another way of registering the location is to register the address of the household. This will of course also lead to a margin of error that has to be assumed.

In the simulations presented in this paper, we have used household density to simulate TV receiver locations. The household density is represented in grids and is not based on the specific household address. 

\subsection{Knowledge Levels}
The goal of this paper is to investigate the potential increase in available spectrum that a CR device can utilize due to TV receiver knowledge. This knowledge is mainly location of the TV receivers, but in addition we define two more knowledge levels that increases the amount of gray space.

Knowledge level 1 (KL1) is defined as \textit{only knowing the TV receivers locations}. Knowledge level 2 (KL2) is defined as \textit{knowing the TV receivers locations and knowing the TV channels each receiver subscribes to}. As different TV channels are multiplexed to increase frequency utilization, KL2 means that a CR device knows which multiplexes (MUXs) a receiver is able to decode.

Knowledge level 3 (KL3) is defined as \textit{knowing the TV receivers locations and which TV channel each TV receiver is currently watching}. This means that a CR device does not only know which MUXs the receiver is able to decode, but actually which MUX the specific receiver is currently watching.

KL2 is a plausible assumption as this information is stored by the service provider. However, this information is business sensitive and not all subscribers would be willing to share this information. Thus obtaining this information requires strict authorization. Given that the privacy issues with KL2 is handled, KL3 is also possible, as only the MUX needs to be reported the TV channel. But it assumes a real-time update from the TV receiver to the database with the current information and real-time update from the database to the CR device. With this knowledge level, the delay between the actual swapping of a TV viewer and the database will be an issue. However, statistics from Germany show that a viewer swaps channels on average 2.3-2.7 times per hour \cite{swapping}.

The main idea behind these knowledge levels is that they increase the level of knowledge of the TV receivers and an analysis of the increase in available spectrum with these knowledge levels are meant to illustrate the basic concept of this paper: the gain from TV receiver information.

\section{CR Device Parameters}
\label{sec:cr-param}
We now estimate the minimum distance between a CR device and a TV receiver, given the specifications in Table \ref{tab:1}. We calculate the distance for two different CR devices: one is a fixed 4 W device thought to represent an IEEE 802.22 base station, the other is a portable 100 mW device.

The required $C/I$ level at the TV receiver will limit the maximum signal strength transmitted by the CR device, denoted $E_{TCR}$. Denoting the maximum allowable signal strength from the CR device at the TV receiver site as $E_{RCR}$, we have
\begin{equation}
L_{\min}[\text{dB}] = E_{TCR}[\text{dB}\mu V/m]-E_{RCR}[\text{dB}\mu V/m]
\end{equation}
where $L_{\min}$ is the minimum required overall path loss between the CR device and the TV receiver.

For a given equivalent isotropically radiated power (EIRP), the equivalent term of field strength must be calculated. Field strength is related to EIRP through the following equation \cite{ntia}:
\begin{equation}
E[\text{dB}\mu V/m] = 10\log_{10}(EIRP\text{[mW]}) - 20\log_{10}(d\text{[m]})+104.8 \nonumber
\end{equation}
where $d$ is a reference distance for the field strength. In this paper we set $d = 1$. 

\subsection{Max CR Field Strength at a TV receiver}
In Table \ref{tab:1} the minimum field strength and protection criterias for the TV receivers were given. We use the values from Ofcom, as they are the most conservative. The minimum field strength of the TV signal at the TV receiver is $E_{RTV} = 50[\text{dB}\mu V/m]$. The requirement for the $C/I$ level for a $8$ MHz band is 33 dB for co-channel and -17 dB for adjacent channel usage.

The maximum allowed CR field strength at the TV receiver is
\begin{equation}
E_{RCR} = E_{RTV} - C/I.
\end{equation}
When substituting the values obtained above, we find the field strength for the co-channel usage to be $E^{co}_{RCR} = 50 - 33 = 17 [\text{dB}\mu V/m]$ and adjacent channel usage to be $E^{ad}_{RCR} = 50 - (-17) = 67 [\text{dB}\mu V/m]$.

\subsection{Propagation Loss Between CR and TV RX}
To calculate the minimum distance between a CR device and a TV receiver we use the Okumura-Hata propagation model. By changing specific parameters of the model one can predict the path loss for three different environments: urban, suburban and rural \cite{antenna}. We choose the pathloss model for the suburban environment, primarily because the areas chosen in the simulations fit this description.

For a suburban environment, the path loss according to the Okumura-Hata model is given by
\begin{align}
L &= 69.55 + 26.16\log_{10}(f_c\text{[MHz]})-13.82\log_{10}(h_b\text{[m]})\nonumber\\
&-a+(44.9-6.55\log_{10}(h_b\text{[m]}))\log_{10}(d\text{[km]})\nonumber\\
&- (2\log_{10}^2(f_c\text{[MHz]}/28)-5.4) \\
a &= (1.1\log_{10}(f_c\text{[MHz]})-0.7)h_m\text{[m]}\nonumber\\
&- (1.56\log_{10}(f_c\text{[MHz]})-0.8)
\end{align}
where L is the pathloss in dB, $f_c$ is the carrier frequency, $h_b$ is the base station height which in this paper correspond to the CR transmitter height in meters, $h_m$ is mobile height which in this paper corresponds to the TV receiver height in meters and $d$ is the distance between the CR transmitter and TV receiver in km.

The minimum distance between a CR device and a TV receiver is given in Table \ref{tab:min-dist}. For the 4 W fixed device we assume a height of 30 m, as this is supposed to illustrate an IEEE 802.22 base station. For the 100 mW device we assume a height of 2 m. The carrier frequency is equal to 650 MHZ and the TV receiver antenna height is set to 10 m. Note that the difference between the minimum required path loss for co-channel and adjacent channel interference is due to the difference between $E^{co}_{RCR}$ and $E^{ad}_{RCR}$.

\section{Simulation Setup}
\label{sec:sim-setup}
To illustrate the advantage of TV receiver registration, we estimate the gray space amount in three municipalities in Norway based on household statistics in these municipalities. Although we have not been able to obtain exact location on the TV receivers in these areas, we have obtained the locations of households in the areas. We therefore make the assumption that each household location represent a TV receiver location. Even though most households owns a TV set, this is an over estimation of the amount of TV receivers in the areas as not all TV sets receives their signal through wireless broadcasts.

The areas chosen are Lillehammer and Vinje. The land area of these municipalities ranges from 478.2 km-3106 km and the population density ranges from 54/km to 1/km. In addition, the population localization is scattered differently for each area. In Vinje the household density is more scattered than in Lillehammer. In Lillehammer the household density is focused in the middle of the area. The population densities of these areas can be seen in Fig. \ref{fig:densities}. Choosing these areas is regarded as appropriate to analyze village-like and rural areas in Norway, where it is considered that cognitive radio and exploitation of the white and gray spaces will have the most impact.

\subsection{TV Broadcast Station Information}
An area may contain households that receive TV signals from different TV stations. By using the coverage planning map by Norways Television [29], the strongest TV transmitter in each area is found. For instance, in Vinje the strongest broadcast station is Rauland, which uses channels 25, 27, 32, 35 and 42. Using information from Norwegian Post and Telecommunications Authority (NPT), there are in addition 12 TV broadcast stations that are connected to Rauland in a single frequency network (SFN) where all broadcast stations use the same channels \cite{npt}. 

In our simulations we assume all households in the different areas receive TV service from either the strongest TV broadcast station or from another station in the same SFN. In reality, there may be exceptions, e.g. if a TV receiver is at the edge of the area it may receive signals from other broadcast stations that are not in the same SFN. However, these exceptions are neglected in this paper.

\subsection{White Space and Gray Space Amount}
In this paper we have assumed that a broadcast station uses 5 channels to provide service to an area. Each of these channels contain a MUX and the channel bandwidth is 8 MHz. The total amount of spectrum used for TV broadcasting in Norway is 320 MHz. White space is defined as the spectrum not occupied by a MUX, nor the channel's adjacent channels. The maximum number of adjacent channels is 10. The white space amount is therefore $320-5\times 8-10\times 8 = 200$ MHz. Gray space is defined as the channels used for broadcasting in the area and those channels adjacent channels. The maximum gray space amount is therefore $5\times 8+10\times 8 = 120$ MHz.

\subsection{Topology and TV Receiver Locations}
\begin{figure*}[t]
\centering
\subfigure[Household locations in Vinje]{
\includegraphics[height = 0.5\columnwidth]{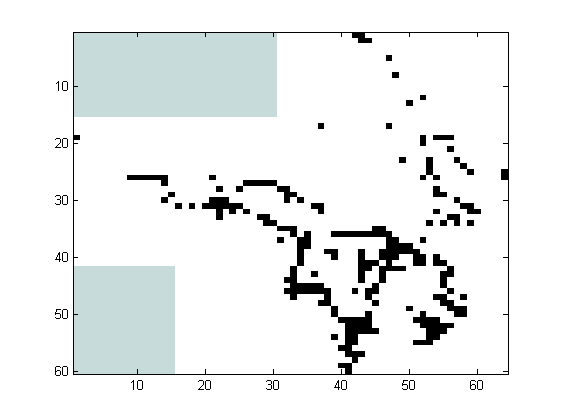} 
}
\subfigure[Household locations in Lillehammer]{
\includegraphics[height = 0.5\columnwidth]{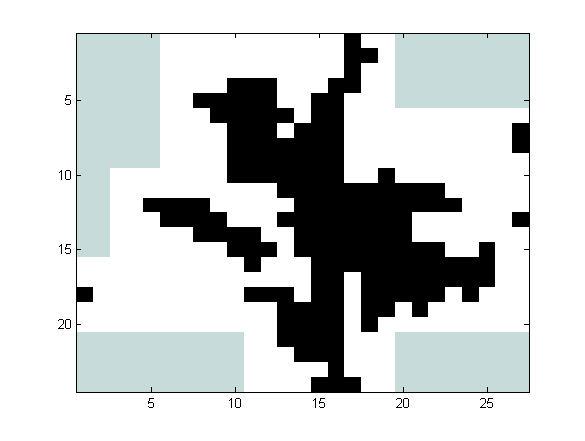} 
}
\caption{Household locations for the municipalities considered in this paper for grid size of 1km.}
\label{fig:densities}
\end{figure*}

The information on household density is retrieved from Statistics Norway (SSB) \cite{ssb}, who provide the statistics for different grid sizes: 100 m and 1 km. For each grid element, the information provides the number of households within each grid. If an element contains a household, it is marked as black as in Fig. \ref{fig:densities}. In the illustration of the densities in Fig. \ref{fig:densities} we have not made an illustration of the number of households in an element, i.e. an element is either white or black.

The method used to map the information from SSB to the grid leads to an overestimation of the area. As described, the area is converted to rectangular $M\times N$ grid, but the original shape of the municipal may be different. For instance, the municipal of Vinje has an area of 3106 km. E.g. for the case where each grid element is 1 km, the information from SSB leads to a grid with size $64\times 60$, which corresponds to an area of 3840 km. This means there are 734 grid elements that are not part of the municipal of Vinje. Neglecting this expansion would lead to an overestimation on the amount of white and gray spaces found in the simulations. We compensate for this expansion by marking a number of grid elements equal to the expansion factor at certain borders of the grid as not being part of the municipal and thus not to be considered as containing gray space. This can be seen in Fig. \ref{fig:densities} where certain parts of the grid for each municipal are crossed out.

\subsection{Protection of the TV Receivers}
Each grid element that contains a TV receiver must be protected from a CR device. In Table \ref{tab:min-dist} we calculated the minimum distance between a CR device and a TV receiver. From this minimum distance we can calculate the minimum number of grid elements surrounding each TV receiver that must be protected, i.e. a CR device in such a grid cannot transmit. 

\begin{figure}[t!]
\centering
\includegraphics[width = 0.9\columnwidth]{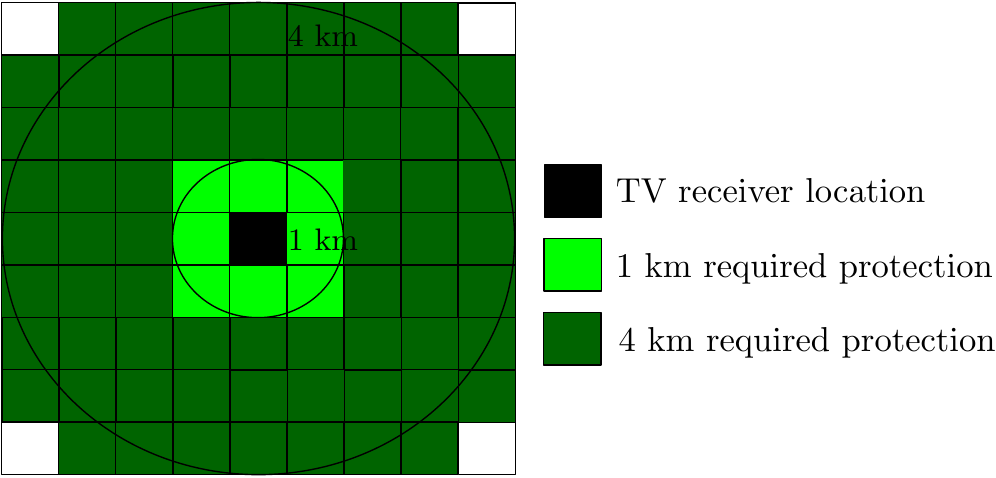}
\caption{Protection degree in distance and corresponding $N$ value}
\label{fig:protection-degree}
\end{figure}

In the simulations we make the assumption that the TV receiver antennas does not have any directional gain. Thus they protected equally in all directions. In Fig. \ref{fig:protection-degree} a minimum protection distance of 1 km and 4 km is shown (where each grid element is 1 km). It is illustrated as circular areas surrounding the TV receiver location, where the TV receiver is illustrated as the black grid element. Since we do not know exactly where the TV receiver is located in the grid element, each grid element that intersects the protected circle must be marked as protected. This leads to a semi-circular protection area of the TV receiver. Given the minimum distance between a CR device and a TV receiver, the minimum distance must be rounded up to the nearest integer consistent with the grid resolution. E.g. for a minimum distance of 7350 m. the protection distance is rounded up to 8 km if we consider a grid resolution of 1 km, and rounded up to 7400 m if we consider a grid resolution of 100 m.

\subsection{Channel Usage for each Knowledge Level}
In the simulations we assume 5 MUXs are transmitted from each TV broadcast station. Thus 5 channels are used for TV broadcasting in a particular area. We denote these MUXs with numbers from 1 to 5. We assume, as is the actual case in many countries, that all TV channels that are freely available to the public are contained in MUX 1. All other TV channels are contained in MUX 2-5.

However, we do make the assumption that each receiver in the topology receives its signal from the same broadcast station, and thus that the 5 channels used by that broadcast station are the only occupied channels. This may not be the case, as many coverage areas overlap. The result is that some TV receivers may not be taken into account in the analysis, thus increasing the gray space. However, this amount is assumed minor compared to the whole topology.

\subsubsection{Knowledge Level 1}
Since KL1 is defined as having no information regarding the actual channels use, all TV receivers within the area are assumed to receive and utilize all 5 MUXs available in the area. As mentioned above, we use household locations to illustrate TV receiver locations. Thus, for KL1 we protect all households from interference from a CR device.

\subsubsection{Knowledge Level 2}
For KL2, we assume the CR device also knows the MUXs a TV receiver subscribes to. Based on data provided by SSB and NTV in Norway \cite{barometer}\cite{digitalkringkasting} we assume that
\begin{itemize}
\item 98\% of all TV receivers are able to use MUX 1.
\item 15\% of all the TV receivers subscribes to MUX 2-5.
\end{itemize}
Although MUX 1 is free, the 2\% that are not able to use MUX 1 are households outside the coverage area of any broadcast station. Note that since a grid element can, and usually does, contain multiple households, the probability that a particular grid element contain a TV receiver that subscribes to MUX 2-5 depends on the number of households in the grid element. E.g. if a grid element contain 20 households the probability that none of the subscribe to MUX 2-5 is small, as on the average 3 households subscribe to MUX 2-5.

\subsubsection{Knowledge Level 3}
It is known that the MUX usage changes with time. This may be due to one MUX containing a particularly popular show for one time interval, or a general factor that people are more likely to use a MUX at particular times. Based on statistics obtained from the market share of viewers and channels in Norway \cite{mediaNorge} we define two time periods: one from 0900-1500h and one from 2000-2300h. The percentage of all TV receivers that utilize the different MUXs in these two time periods are shown in Table \ref{tab:kl3-percent}. Note that the sum of these percentages is the average percent of people watching TV sometime in these time periods. I.e. in time period 1, 10\% of the TV receivers are utilized, while in time period 2, 45\% of the TV receivers are utilized.

One may also note that the sum of the market share for MUX 2-5 in time period 2 is 25.8\%, which may seem to contradict the assumption that only 15\% of the TV receivers subscribe to MUX 2-5. However, this is the result for all TV viewings, also those that obtain their signal from cable or satellite. Thus, for instance MUX 2s market share of 12.7\% means that of those 15\% subscribing to MUX 2-5, 12.7\% is actually watching a TV channel contained in MUX 2 in this time period.

\section{Simulation Results}
\label{sec:sim-res}
\begin{figure}[t!]
\centering
\subfigure[1 km]{
\includegraphics[height = 0.5\columnwidth]{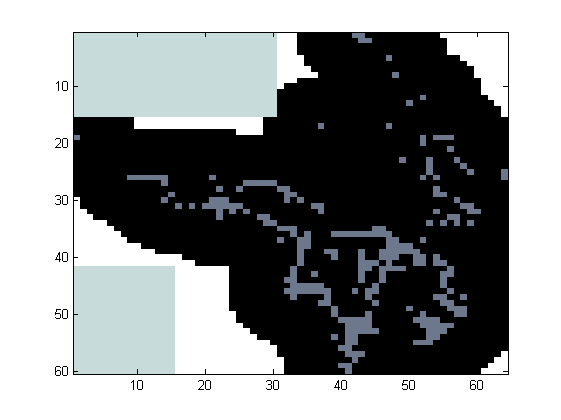} 
}
\subfigure[100 m]{
\includegraphics[height = 0.5\columnwidth]{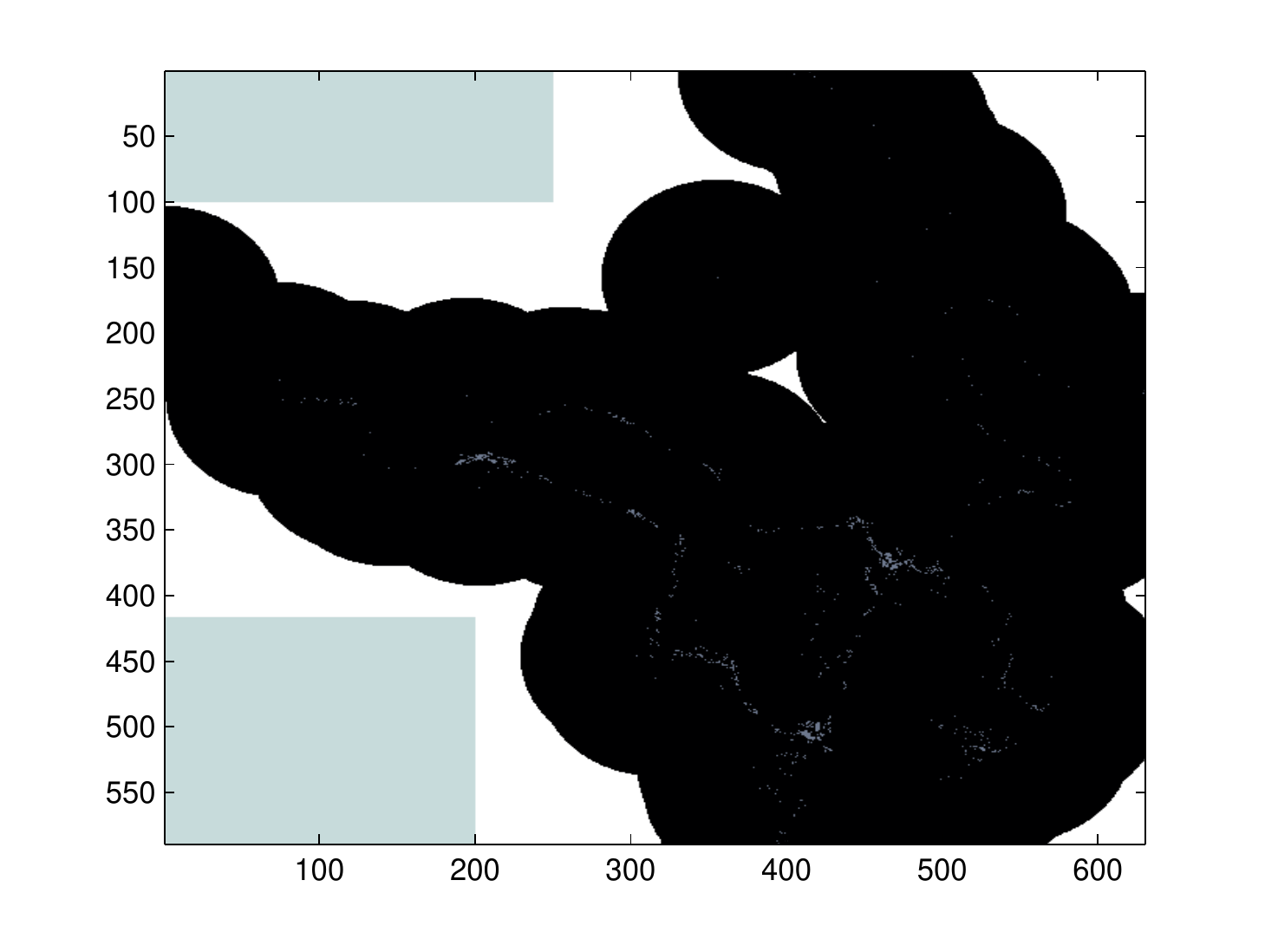} 
}
\caption{Areas where co-channels can be utilized (white areas), for (a) 1 km resolution and (b) 100 m resolution.}
\label{fig:vinje-diff-100-1000}
\end{figure}

We now provide the simulation results from the municipalities. Fig. \ref{fig:vinje-diff-100-1000} shows the difference between using 1 km resolution and 100 m resolution for co-channel usage. The rest of the results are given in plots where the y-axis is the amount of gray space (in MHz) and the x-axis gives the percentage of the area the can utilize it. E.g. in Vinje, with KL1, 100 m grid size and a transmit power of 4 W EIRP, 50\% of the area contain 80 MHz of gray space.

\begin{figure}[t!]
\centering
\subfigure[Knowledge level 1, Vinje]{
\includegraphics[height = 0.3\columnwidth]{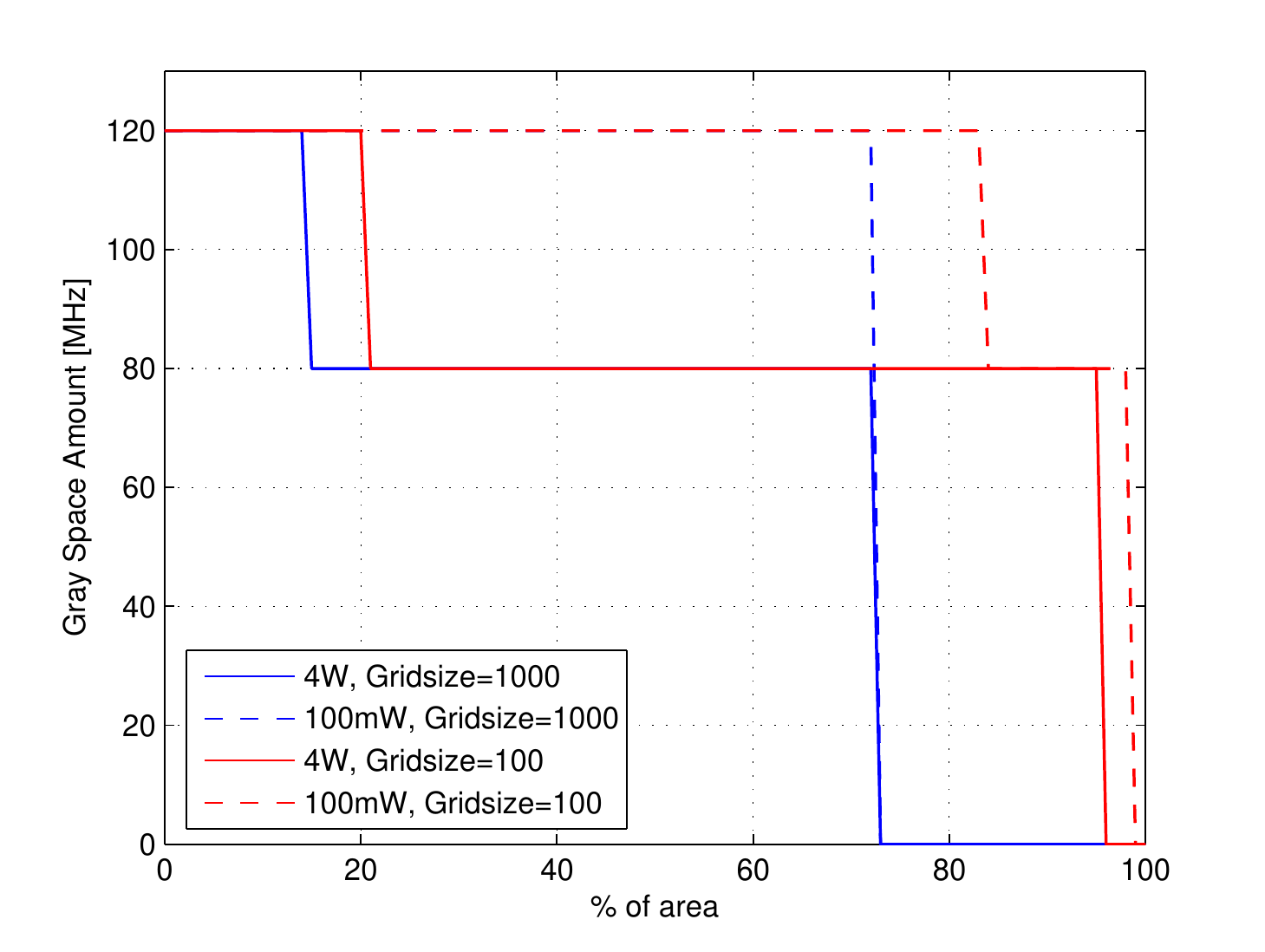} 
}
\subfigure[Knowledge level 2, Vinje]{
\includegraphics[height = 0.3\columnwidth]{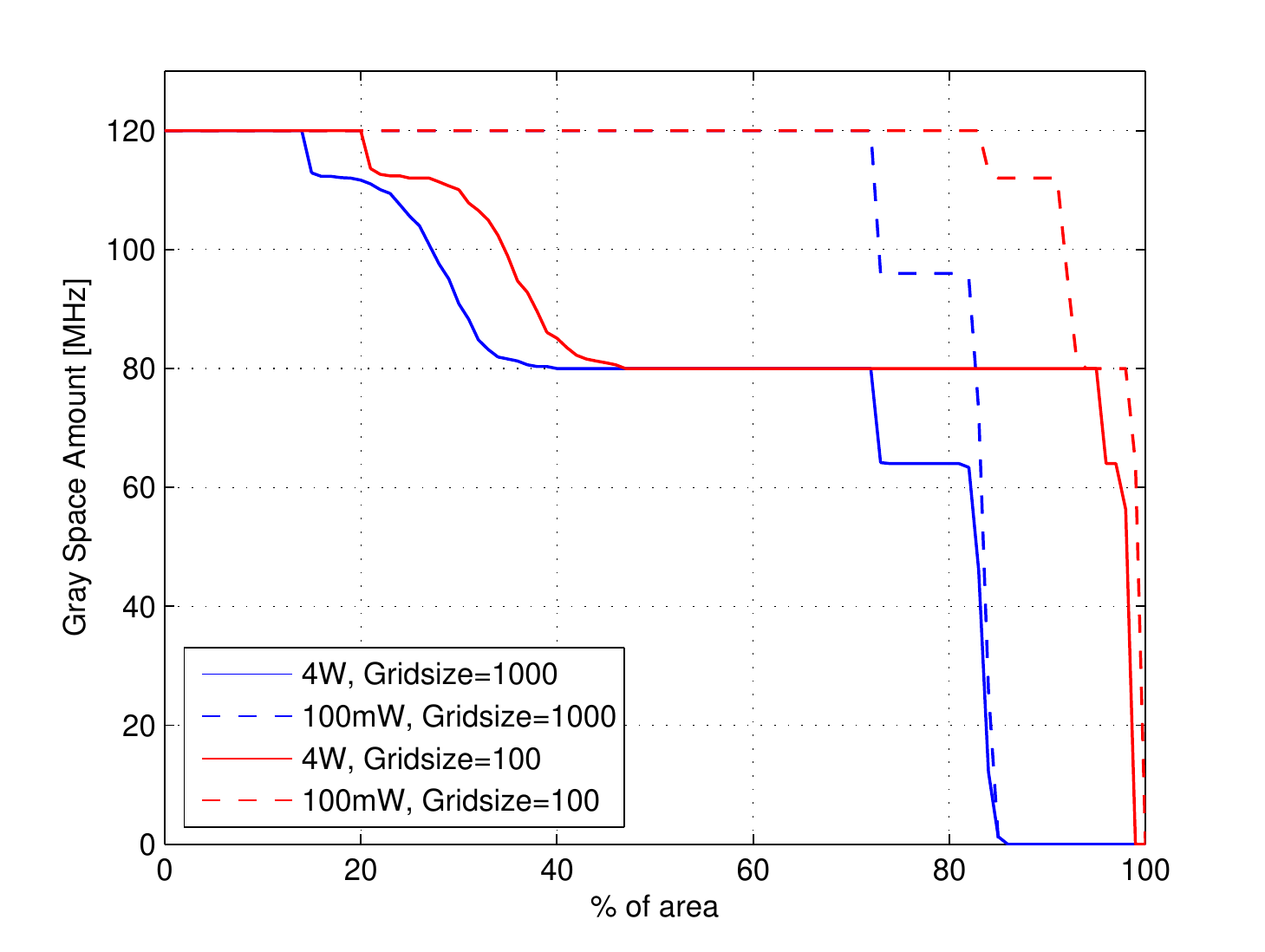} 
}
\subfigure[Knowledge level 3, time period 1, Vinje]{
\includegraphics[height = 0.3\columnwidth]{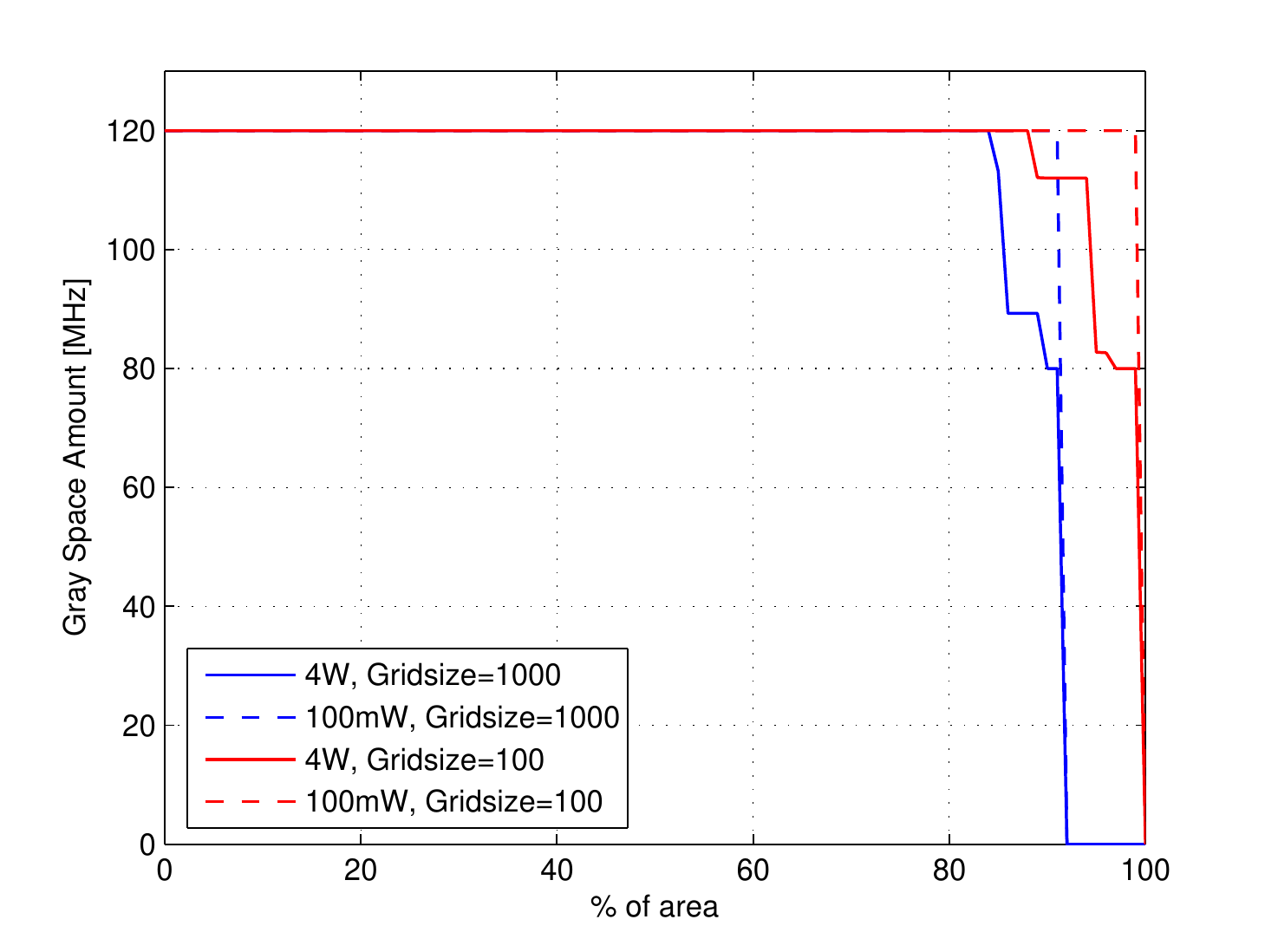} 
}
\subfigure[Knowledge level 3, time period 2, Vinje]{
\includegraphics[height = 0.3\columnwidth]{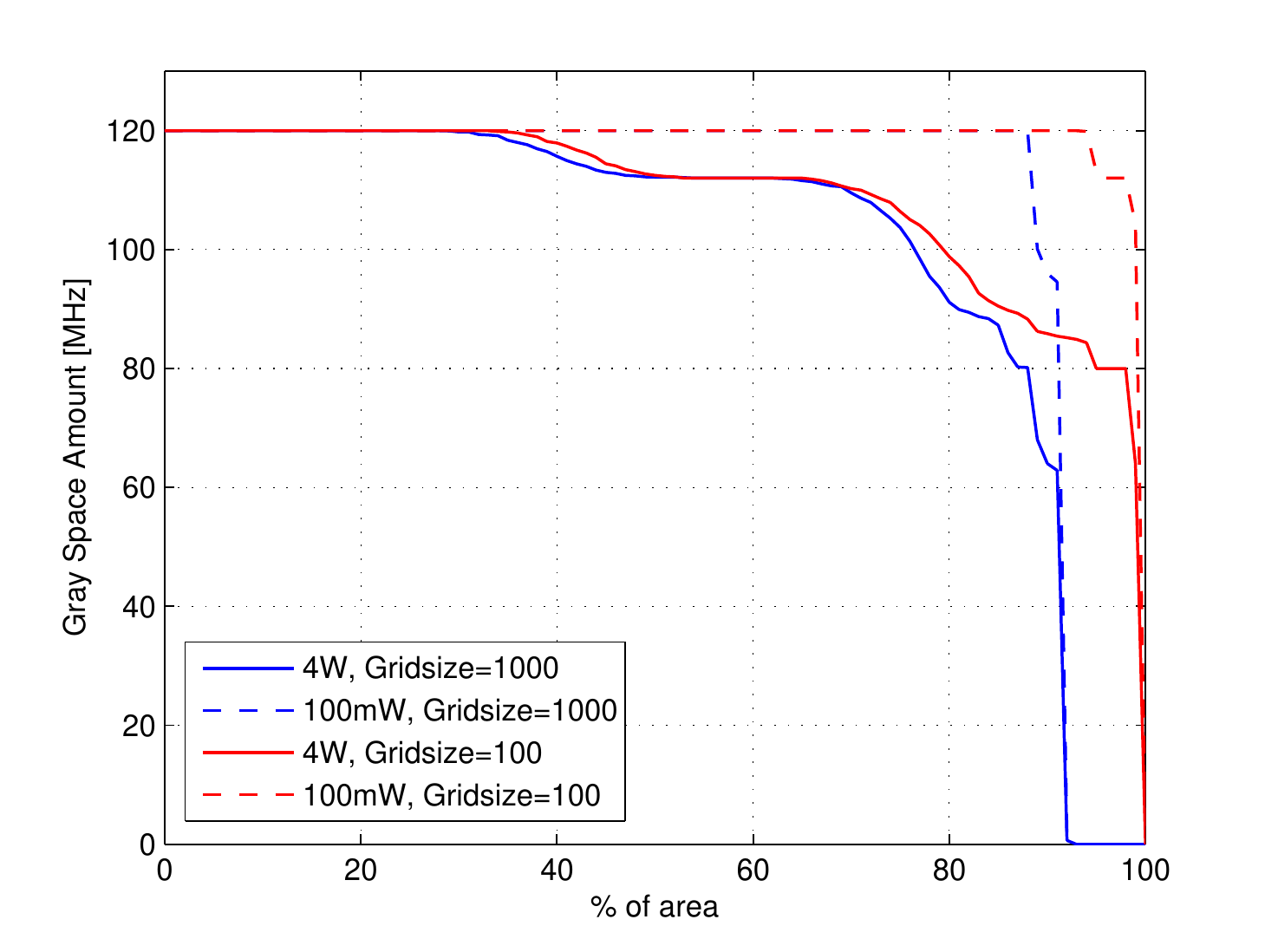} 
}
\caption{Amount of gray space in Vinje for different knowledge levels and grid resolution.}
\label{fig:res-gray-space-vinje}
\end{figure}

Fig. \ref{fig:res-gray-space-vinje} shows the amount of grays space in Vinje for the different knowledge levels and time periods. Our expecations are to a large extent verified: increasing grid resolution increases gray space and increasing the information about the TV receivers increases gray space. We also see that curves for grid resolution of 100 m mimics that 1 km with an off-set of some percentage in exploitable areas. E.g. in Fig. \ref{fig:res-gray-space-vinje}(a) the plot for a 4 W EIRP device with 1 km grid size drops from 120 MHz to 80 MHz at 15\% of the area, whereas with 100 m grid size the drop occur at 20\%. The same occur for 100 mW EIRP, only at a difference of 10\% in area. However, we see that the percentage of area that contain 80 MHz of gray space is much larger for 100 m resolution than for 1 km resolution, as with 1 km no more than 70\% of the area contain any gray space even for 100 mW EIRP.

With increased information the gap between 100 m and 1 km resolution decreases. The plots are also continuous even though the amount of gray space is discrete (e.g. 120, 96, 80, 64). This is due to the fact that for a particular instance of simulation the amount is discrete, but since the amount depends on the locations chosen to be TV receivers, we have averaged the results over 100 realizations and thus obtained semi-continuous plots. 

\begin{figure}[t!]
\centering
\subfigure[Knowledge level 1, Lillehammer]{
\includegraphics[height = 0.3\columnwidth]{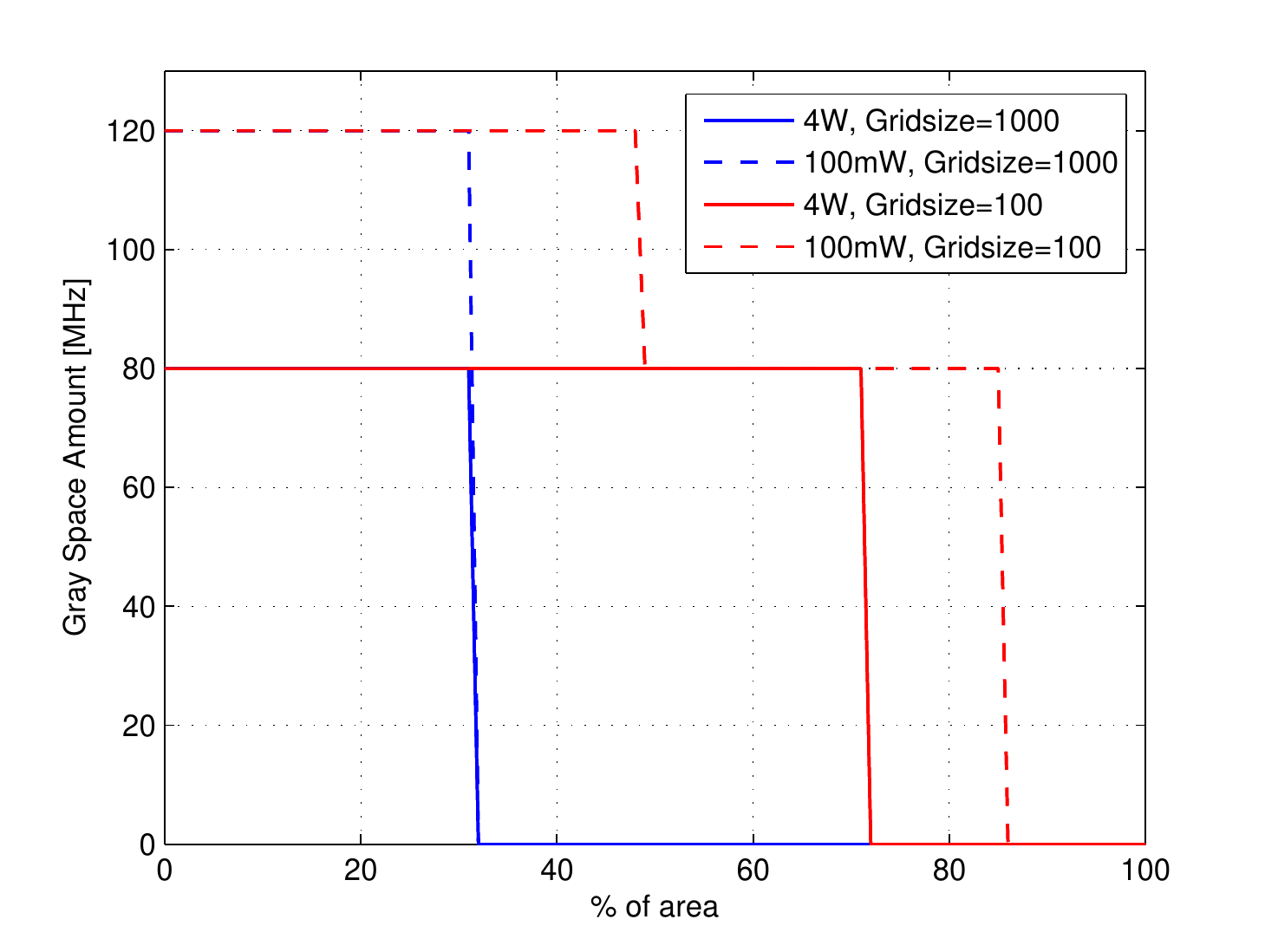} 
}
\subfigure[Knowledge level 2, Lillehammer]{
\includegraphics[height = 0.3\columnwidth]{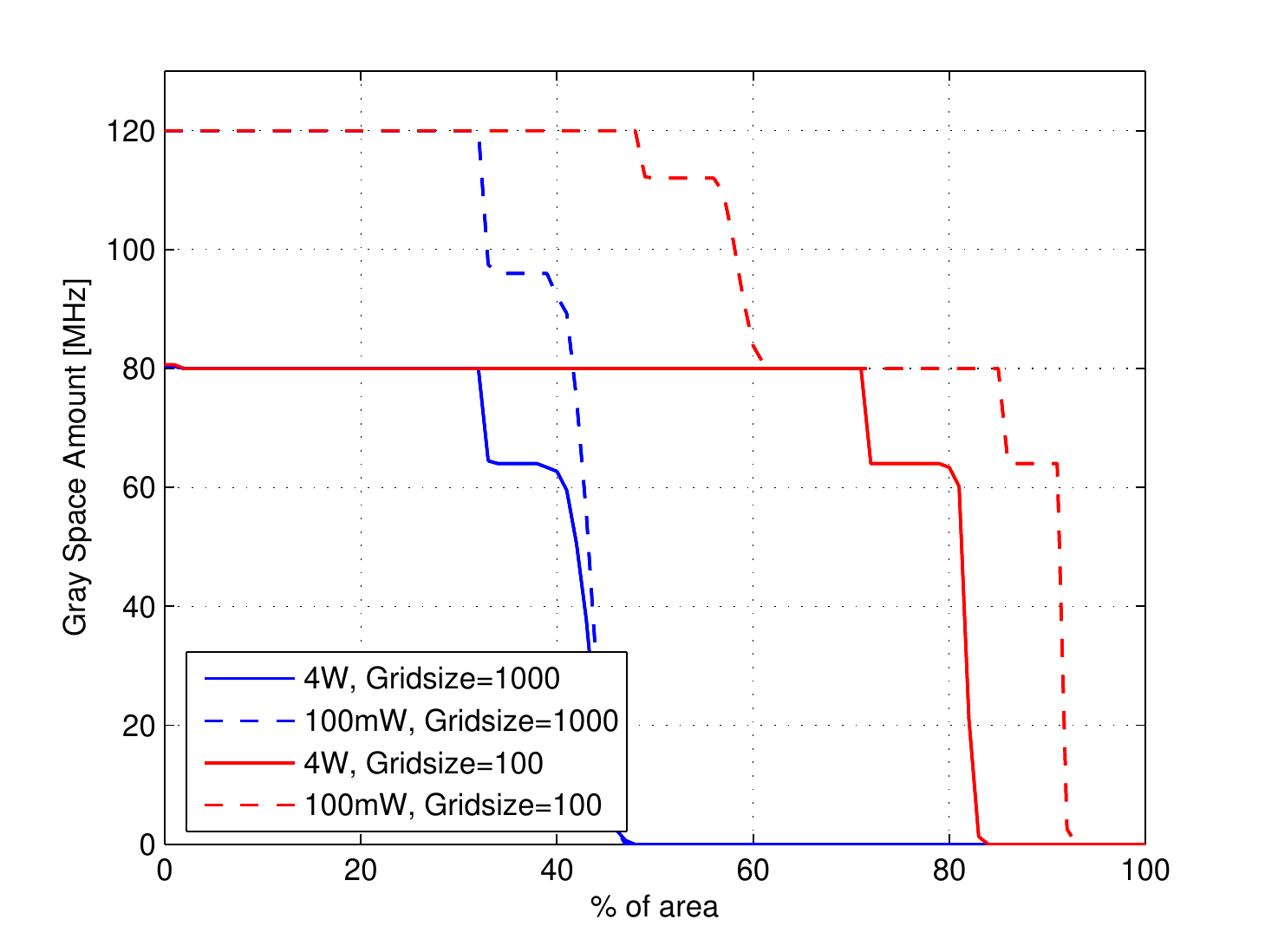} 
}
\subfigure[Knowledge level 3, time period 1, Lillehammer]{
\includegraphics[height = 0.3\columnwidth]{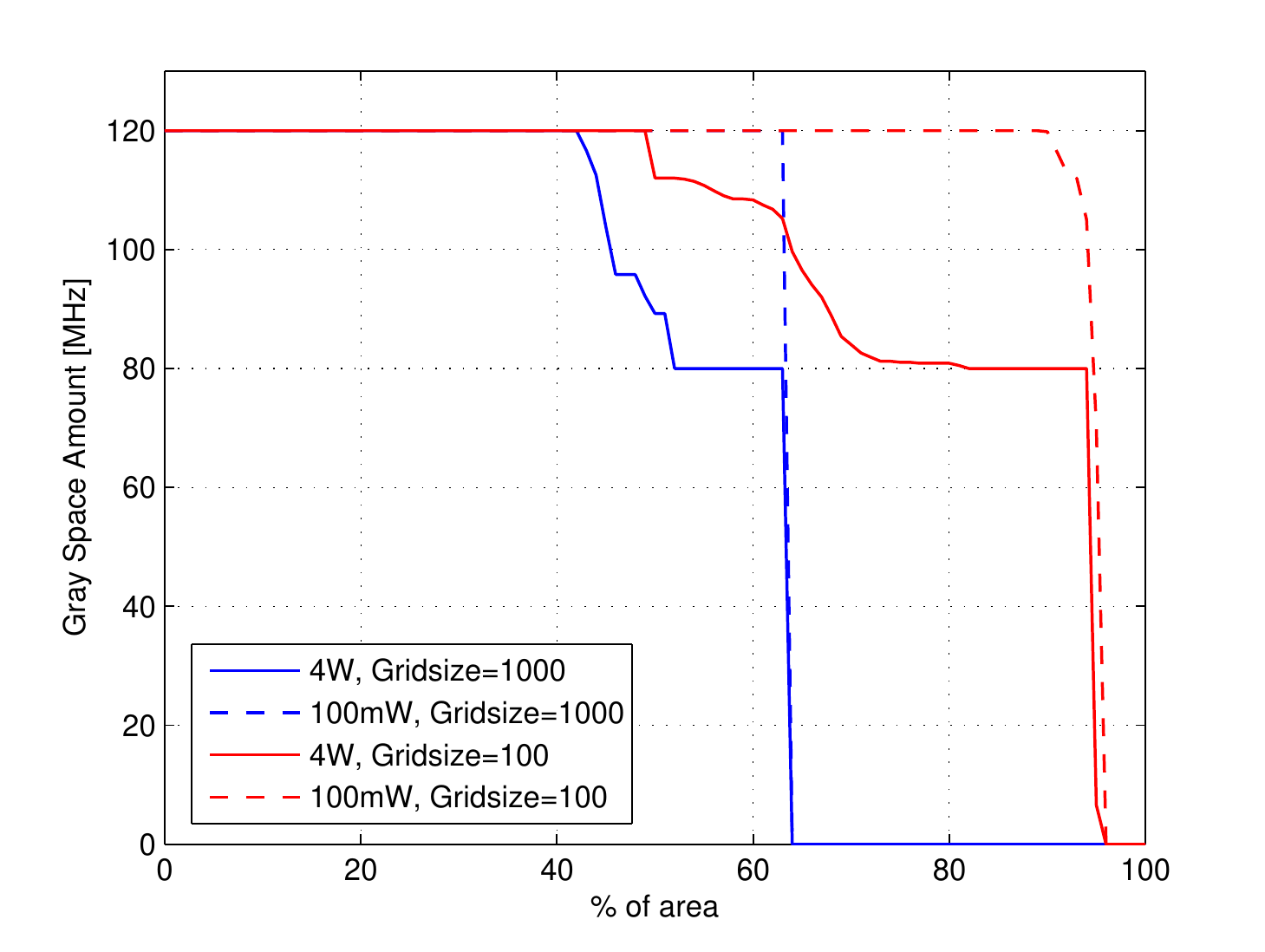} 
}
\subfigure[Knowledge level 3, time period 2, Lillehammer]{
\includegraphics[height = 0.3\columnwidth]{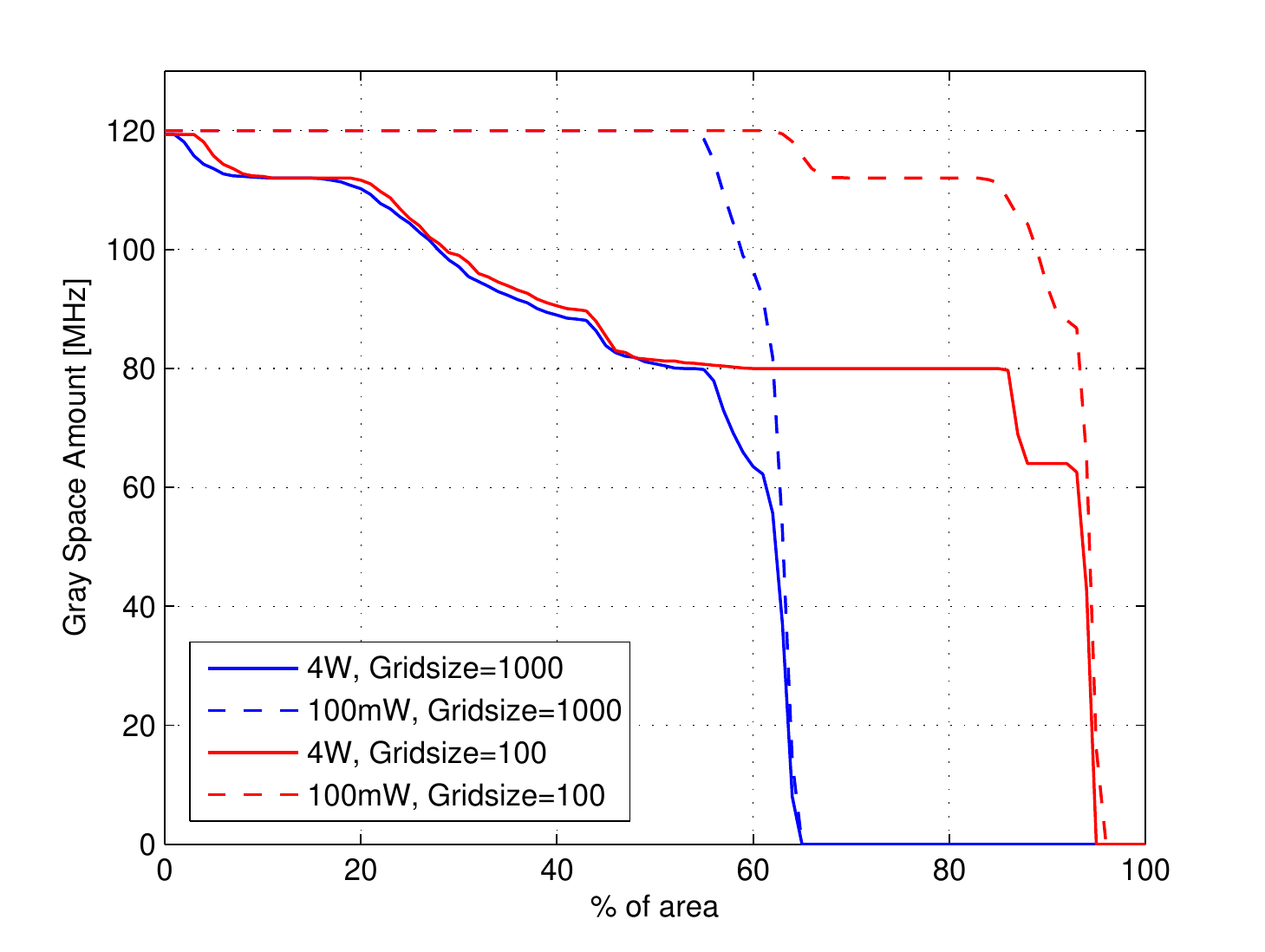} 
}
\caption{Amount of gray space in Lillehammer for different knowledge levels and grid resolution.}
\label{fig:gray-space-lillehammer}
\end{figure}

Fig. \ref{fig:gray-space-lillehammer} shows the same results for Lillehammer. These results show a larger gap between 100 m and 1 km grid resolution. This is due to Lillehammer being a more densely populated municipality than Vinje.

In Table \ref{tab:res-house} the number of households that can utilize the gray space is given. This is an important aspect of the results, since even though 50\% of the area may contain grays space, this is of no value if no one can utilize it. However, note that our statistics does not reveal the location of business centers and other areas that may find additional spectrum valuable.

\section{Discussion}
\label{sec:discussion}
The hypothesis of this paper is that knowledge of the TV receivers will increase the spectrum available for a CR device. The simulations in the previous section verifies the hypothesis to a large degree. We now discuss the results obtained in this paper.

\subsection{Knowledge Levels and Grid Size}
Depending on the CR device considered, knowledge of the TV receiver locations can increase the available spectrum by 80-120 MHz in certain locations. Although we have used household statistics to locate TV receivers in different municipalities in Norway (which we consider a good representation of the actual TV receiver locations, since most households are concentrated in groups with high density), only knowing the TV receiver locations does not enable any households to utilize the available spectrum. This may seem to limit the applicability of the gray space found with this knowledge level, but we have not taken into account type of locations that may be interested in additional spectrum. For instance, our statistics does not show where business areas are located, which are potential users interested in gray space.

Knowledge level 2 leads to an increase of 10\% in areas that can exploit grays space compared to KL1. The reason for the this marginal increase is that although only 15\% of TV receivers are assumed to subscribe to MUX 2-5, a grid element may contain multiple TV receivers. E.g. if a grid element contain 10 TV receivers the probability of none of them subscribing to MUX 2-5 is $\sim 0.2$. One of the advantages of using a smaller grid size is thus that the probability that a TV receiver within a grid element subscribes to MUX 2-5 is lower than with bigger grid sizes. The main argument for why KL2 is worth while implementing is given in Table \ref{tab:res-house}. While KL1 does not allow any households to utilize gray space, depending on the device considered and grid size, KL2 allows between 117 and 669 households to utilize gray space.

Knowledge level 3 leads to the greatest increase in grays space, especially for time period 1. This is not surprising as KL3 is defined as having real time information on TV viewings. We also see from the results in Fig. \ref{fig:res-gray-space-vinje} and Fig. \ref{fig:gray-space-lillehammer} that difference in gray space for the two grid sizes decreases with increased information about the TV receivers. For KL3 in Vinje the difference between 1 km grid size and 100 m grid size in terms of gray space amount is almost neglectable. However, for Lillehammer the gap is still large, which is due to Lillehammer having a higher population density than Vinje.

The difference between using 1 km grid size as opposed to 100 m grid size can also be seen in the number of households that can utilize the gray space, as is given in Table \ref{tab:res-house}. Although the difference between using 1 km grid size and 100 m grid size in terms of gray space amount in Vinje is almost neglectable, the difference in number of households that can utilize that gray space is not. For instance, for a 4 W EIRP CR device, the number of householsd that can utilize 24-64 MHz of gray space is 110 for KL2 and 1 km grid size. For the same parameters and 100 m grid size the number is 377.

\subsection{Business Aspect}
The main argument for opening the white spaces for public use is that people need spectrum and that the spectrum used for TV broadcasting is severely under utilized. Therefore, let those who need spectrum use it. For instance, in the US, if one satisfies the criteria given by the FCC, anyone can transmit in the white spaces. In other countries this is not the case; one can only transmit in the white spaces with specific permission by the license holder.

Either way, the results found in this paper has value to both the license holder and a service provider. The license holder may be interested to know how ineffective the utilization of the frequencies is at certain locations so that the license holder can provide additional or other services to its customers. The license holder can also lease out the gray spaces (and in some countries the white spaces) to other service providers. The criteria is that the TV receivers must not be disturbed. As the profit will depend on how much spectrum the license holder is able to provide, providing as much knowledge about the TV receivers as possible will benefit the license holder.

For a service provider in need of spectrum, the amount that he can obtain and the location where the spectrum can be utilized have value. If no one can utilize the spectrum, the service provider will not make a profit. To clearly estimate the potential of gray space for a service provider a more thorough investigation of household locations and business locations are necessary.

\subsection{Assumptions}
The main assumptions in this paper is that the minimum distance can be estimated based on the propagation path loss from the Okumura-Hata model, that households represent TV receiver locations and that all antennas are omni-directional. We will now discuss how these assumptions may have affected the results.
\subsubsection{Okumura-Hata Model}
Consider a 4 W EIRP CR device and 1 km grid size. In this case the minimum protection distance is 8 km for co-channel and 1 km for adjacent channel usage. Using the Okumura-Hata model the setup used in this paper, 8 km correspond to a pathloss of 125 dB. This gives a carrier-to-interference ratio of
\begin{equation}
C/I = 50 \text{[dB}\mu V/m\text{]} - 140.8 \text{[dB}\mu V/m\text{]} + 125 \text{[dB]} = 34.2 \text{[dB]} \nonumber.
\end{equation}
The required $C/I$ level defined by Ofcom is 33 dB, and our protection of the TV receivers gives only an error margin 1.2 dB, which might be to low. If we on the other hand consider adjacent channel interference, the protection distance was set to 1 km for 1 km grid sizes. 
This gives a carrier-to-interference ratio of
\begin{equation}
C/I = 50 \text{[dB}\mu V/m\text{]} - 140.8 \text{[dB}\mu V/m\text{]} + 93.2 \text{[dB]} = 2.4 \text{[dB]} \nonumber,
\end{equation}
whereas the $C/I$ level defined by Ofcom is -17 dB. This gives an error margin of 19.4 dB which is considered sufficient. For 100 m grid sizes the error margin reduces to 0.8 dB for co-channel interference and 1 dB for adjacent channel interference. 

%
%

\subsubsection{TV Receiver Location}
In the simulations based on statistics in Norway we used household statistics to locate the TV receivers. As most TV receivers are located in households, this is a valid assumption. However, not all households obtain their TV service from TV broadcast stations, a fair amount receive their TV signal from ground cable or satellite. Finding TV receivers from household statistics is therefore an overestimation of the number of TV receivers in an area. One can therefore make the assumption that representing TV receiver locations based on household statistics in a rural area, in fact represents the real distribution of TV receivers in a suburban area. Thus although our simulations were done in two suburban municipalities, the results might be more applicable to small to medium sized cities.

\subsubsection{Omni-directional Antennas}
By assuming omni-directional antennas we assume a worst case scenario where CR transmission achieves the same antenna gain as the TV signal at the TV receiver. If the CR device has a directional antenna and the desired receiver of the CR transmission is located away from the TV receiver, the minimum distance required between the CR transmitter and TV receiver can be increased.

In addition, most TV receiver antennas have a directional antenna gain of more than 4 dB. Thus, the computed minimum distance is a worst case estimate for when a CR device emission is parallel to the TV transmission.

Finally one should note that in this paper we have only located where it is possible for a CR device to operate without creating harmful interference to the TV receivers. We have not considered the areas where the TV broadcast station will create to much interference for a CR device to operate. This aspect of pollution of the white spaces is discussed in \cite{harrison} where it is argued that this pollution reduces the amount white spaces. As gray spaces are found inside the service area of a TV broadcast station, this is major issue with regard to the amount that can be utilized.

\section{Conclusion}
\label{sec:conclusion}
In this paper we have investigated the possibility of a cognitive radio device exploiting a TV broadcast stations spectrum within its service area. Through registration of TV receiver location we have shown that the available spectrum can increase by as much as 120 MHz compared to only knowing the TV broadcast station location. In addition to knowledge of TV receiver locations we introduced two more knowledge levels; knowledge about the TV receivers subscription and real time viewing knowledge. For these knowledge levels we simulated TV receiver locations in Norway based on household statistics for two different TV receiver location resolution; 1 km and 100 m.

From our results we show that in a sparsely population area, 100 m grid size as opposed to 1 km leads to an increase in spectrum of 10-15\%, while for more populated areas the difference can be as much as a factor of 2. For the two municipalities considered in this paper, we saw that between 30-70\% of the areas could utilize an additional 80 MHz by TV receiver registration. By increased information and higher resolution this could be increased to occur at 70-95\% of the areas.

As Ofcoms and FCCs rules for cognitive radio activity in theory opens for exploitation of a TV broadcast stations spectrum within its service area, our study also confirms that the TV white spaces in general should be opened for cognitive radio activity and that both Ofcoms and FCCs rules for cognitive radio operation are conservative.

\bibliographystyle{ieeetr}
\bibliography{ref,bib}{}
\clearpage
\begin{table*}[t!]
\centering
\caption{Technical parameters defined by Ofcom and FCC}
\begin{tabular}{|l|l|l|}
\hline
\textbf{Parameters} & \textbf{Ofcom/UK} & \textbf{FCC/USA} \\
\hline
Bandwidth [MHz] & 8 & 6 \\
\hline
Min. field strength for service & 50 dB$\mu V/m$ & 41 dB$\mu V/m$ \\
\hline
Receiver height & 10 m & 10 m \\
\hline
Location accuracy & 100 m  & 50 m \\
\hline 
Co-channel interference & 33 dB & 23 dB \\
Power limit & As specified by the database & Fixed device: 4 W \\
\hline
Adjacent channel interference & -17 dB & -26/-28 dB \\
Power limit & 50 mW & Portable device: 40/100 mW \\
\hline
\end{tabular}
\label{tab:1}
\end{table*}

\begin{table*}
\centering
\caption{Minimum distance between a CR device and a TV receiver with 4 W EIRP}
\begin{tabular}{|l|l|l|l|l|}
\hline
 & \textbf{Channel Type} & \textbf{CR field strength} & \textbf{Minimum required} & \textbf{Min. distance} \\
 & & & \textbf{path loss L[dB]} & \\
\hline
\multirow{2}{*}{\textbf{4 W EIRP}} & co-channel & 140.8 & 123.8 & 7.35 km \\
& adjacent channel& & 73.8 & 281 m \\
\hline
\multirow{2}{*}{\textbf{100 mW EIRP}} & co-channel & 124.8 & 107.8 & 910 m \\
& adjacent channel & & 57.8 & 62 m\\
\hline
\end{tabular}
\label{tab:min-dist}
\end{table*}

\begin{table}[t]
\centering
\caption{Percentage of market share for the two time periods}
\begin{tabular}{|l|l|l|}
\hline
\textbf{MUX} & \textbf{Time period 1 (9AM-3PM)} & \textbf{Time period 2 (8PM-11PM} \\
\hline
MUX 1 & 3.46\% & 19.3\%\\
MUX 2 & 2.45\% & 12.7\% \\
MUX 3 & 1.59\% & 6.2\%\\
MUX 4 & 1.78\% & 5.7\% \\
MUX 5 & 0.7\% & 1.2\% \\
\hline
\end{tabular}
\label{tab:kl3-percent}
\end{table}

\begin{table*}[t!]
\centering
\caption{Number of households that can utilize the gray space in Vinje}
\begin{tabular}{|l|l|l|l|l|l|}
\hline
Device & Grid Resolution & Gray space amount & KL2 & KL3, T1 & KL3, T2 \\
\hline
\multirow{6}{*}{4 W} & \multirow{3}{*}{1 km} & 24-64 MHz & 110 & 61 & 178 \\
& & 72-96 MHz & 6 & 75 & 518 \\
& & 96 MHz $<$ & 0 & 868 & 308 \\
\cline{2-6}
& \multirow{3}{*}{100 m} & 24-64 MHz & 377 & 0 & 257 \\
& & 72-96 MHz & 8 & 109 & 737 \\
& & 96 MHz $<$ & 0 & 1221 & 335 \\
\hline
\multirow{6}{*}{100 mW} & \multirow{3}{*}{1 km} & 24-64 MHz & 0 & 24 & 25 \\
& & 72-96 MHz & 114 & 79 & 490 \\
& & 96 MHz $<$ & 1 & 901 & 489 \\
\cline{2-6}
& \multirow{3}{*}{100 m} & 24-64 MHz & 511 & 0 & 45 \\
& & 72-96 MHz & 155 & 29 & 482 \\
& & 96 MHz $<$ & 3 & 1301 & 802 \\
\hline
\end{tabular}
\label{tab:res-house}
\end{table*}

\end{document}